\documentclass[twocolumn,english,pra,amsmath,floatfix,showpacs,superscriptaddress,footinbib]{revtex4}
\usepackage{lmodern}
\usepackage[utf8]{inputenc}
\usepackage{color}
\usepackage{babel}
\usepackage{amsmath}
\usepackage{amssymb}
\usepackage{graphicx}
\usepackage[colorlinks=true,linkcolor=blue,citecolor=blue,
    urlcolor=blue]{hyperref}
\usepackage{epstopdf}
\usepackage{etoolbox}
\apptocmd{\sloppy}{\hbadness 10000\relax}{}{}

\newcommand{\ket}[1]{| #1 \rangle} 
\newcommand{\bra}[1]{\langle #1 |} 
\newcommand{\up}[0]{\uparrow} 
\newcommand{\dn}[0]{\downarrow} 
\newcommand{\mean}[1]{\langle #1 \rangle} 
\newcommand{\SiSj}[2]{\mathbf{S}_{#1}\cdot\mathbf{S}_{#2}}
\newcommand{\mSiSj}[2]{\mean{\SiSj{#1}{#2}}}

\begin{document}

\bibliographystyle{apsrev}

\title{Thermal entanglement in a triple quantum dot system}

\author{M. Urbaniak}
\affiliation{Institute of Molecular Physics, Polish Academy of Sciences,
ul. M. Smoluchowskiego 17, 60-179 Pozna{ń}, Poland}
\author{S. B. Tooski}
\affiliation{Institute of Molecular Physics, Polish Academy of Sciences,
ul. M. Smoluchowskiego 17, 60-179 Pozna{ń}, Poland}
\affiliation{Faculty of Mathematics and Physics, University of Ljubljana,
 and Jozef Stefan Institute, Ljubljana, Slovenia}
\author{A. Ram{š}ak}
\affiliation{Faculty of Mathematics and Physics, University of Ljubljana,
 and Jozef Stefan Institute, Ljubljana, Slovenia}
\author{B. R. Bu{ł}ka}
\affiliation{Institute of Molecular Physics, Polish Academy of Sciences,
ul. M. Smoluchowskiego 17, 60-179 Pozna{ń}, Poland}

\begin{abstract}
We present studies of thermal entanglement of a three-spin system in triangular symmetry. Spin correlations are described within an effective Heisenberg Hamiltonian, derived from the Hubbard Hamiltonian, with super-exchange couplings modulated by an effective electric field. Additionally a homogenous magnetic field is applied to completely break the degeneracy of the system.  We show that entanglement is generated in the subspace of doublet states with different pairwise spin correlations for the ground and excited states. At low temperatures thermal mixing between the doublets with the same spin  destroys entanglement, however one can observe its restoration at higher temperatures due to the mixing of the states with an opposite spin orientation or with quadruplets (unentangled states) always destroys entanglement. Pairwise entanglement is quantified using concurrence for which analytical formulae are derived in various thermal mixing scenarios. The electric field plays a specific role -- it breaks the symmetry of the system and changes spin correlations. Rotating the electric field can create maximally entangled qubit pairs together with a separate spin (monogamy) that survives in a relatively wide temperature range providing robust pairwise entanglement generation at elevated temperatures.
\end{abstract}


\maketitle

\section{Introduction}

Entanglement is recognized to be a key resource in quantum information processing tasks such as quantum computation, teleportation and cryptography \cite{Nielsen}. In recent years the two-qubit entanglement which includes variety of interactions have been studied extensively \cite{Horodecki}. However, the three-qubit entanglement states have been revealed to hold advantage over the two-qubit states in quantum teleportation \cite{Karlsson}, dense coding \cite{Hao} and quantum cloning \cite{Bru}. DiVincenzo {\it et al.} \cite{DiVincenzo2000} proposed quantum computations scheme to measure and control states by logical gates in a three-qubit spin system with exchange interactions in quantum dots, in which logical qubits are encoded in the doublet subspace (see also \cite{Weinstein05,Bacon03}). This scheme uses the advantage of the decoherence-free subspace (DFS) theory in which quantum information encoded over the entangled subspace of system states is robust against decoherence and error processes \cite{Zanardi97,Zanardi98,Fong11,Lidar1998,Kempe}. Recently experimental efforts have been undertaken to fabricate such triple quantum dot structures and to perform coherent spin manipulations \cite{Takakura10,Gaudreau,Laird10}.

Fast reliable spin manipulation in quantum dots is one of the most important challenges in spintronics and semiconductor-based quantum information. Recent experiments show that this goal can be better achieved in electrically gated quantum dot
qubits \cite{Petta2005,Shi}. There have been ongoing theoretical studies to manipulate spin properties of triple quantum dot systems \cite{Li07,Hsieh2010,Bulka11,Hsieh2012} and to apply them in the quantum information precessing controlled by external electric field \cite{Shi,Kuba12}. For the quantum computing most relevant is the ground-state entanglement and its relation with the spin-spin correlation functions and chirality. However, in realistic systems and for potential application to quantum information processing, it is crucial to understand also the entanglement stability at elevated temperatures, which is one of the main goals of this paper.

One of the significant goals for quantum computing and quantum communication is to find an entangled source in solid state systems at a finite temperature. The problem of entanglement between two qubits interacting via the Heisenberg interaction at a nonzero temperature was investigated by Nielsen \cite{Nielsen2000}. A similar problem of the variation of entanglement with temperature and magnetic field in a one dimension finite Heisenberg chain were studied by Arnesen {\it et al.} \cite{Arnesen}, and they used for the first time the notion of thermal entanglement. Thereafter, the issue of entanglement in thermal equilibrium states has been the subject of a number of papers dealing with different aspects of the problem \cite{Ramsak09,Ramsak07,Ramsak06,Zhang07,Huang09}.

Recently experimental efforts have been undertaken to fabricate such triple quantum dot (TQD) structures \cite{Takakura10,Gaudreau,Laird10} and to perform coherent spin manipulations according to the scheme proposed by DiVincenzo {\it et al} \cite{DiVincenzo2000}. It was shown that the quantum states of a coded qubit in a TQD indeed can be manipulated by tuning the gate voltages. Laird {\it et al.} \cite{Laird10} demonstrated experimentally the initialization, coherent exchange, and readout of a coded qubit based on TQD with three electrons by adopting a specific pulsing technique \cite{Petta2005} that was first exploited to demonstrate coherent exchange and readout in a double dot system. Also TQD system with a triangular symmetry to observe spin frustration \cite{Andergassen} was created and a state of single electron spin using single shot readout technique \cite{Morello} was measured. Measuring single spins in many spins systems is the next goal that would allow to check results presented in this paper.
To our knowledge, although several studies in the literature describe the entanglement in a three-qubit system \cite{Huang09}, the thermal entanglement manipulated by an external electric field has not been considered so far. Here we present both analytical and numerical analysis of the thermal entanglement of triple quantum dots as a three-spin qubit described by an effective Heisenberg Hamiltonian at thermal equilibrium in the presence of both external magnetic and electric fields whose magnitude and direction provide additional degrees of freedom to control the entanglement. Since there is no unique measure of entanglement for tripartite entanglement \cite{Dur2000,Brandao2008}, we have calculated monotone concurrence \cite{Hill97} as a measure of entanglement between two qubits of the three-qubit spin system. Then we study the combined influence of magnetic field strength and direction of electric field on entanglement of the system.

We also determine critical temperature $T_0$ for a chosen set of parameters of the system, beyond which concurrence vanishes in agreement with Ref.\cite{Fine2005}. However, the concurrence can be a nonmonotonic function of temperature: at low temperatures regime the thermal mixing destroys the entanglement,  while at higher temperatures one still can observe partial restoration of the entanglement. The studies add a new useful tool for manipulating the entanglement of the three-qubit Heisenberg model.

\section{Model of three spins in a quantum dot system}

\begin{figure}[t]
\centering
\includegraphics[clip,width=0.3\textwidth]{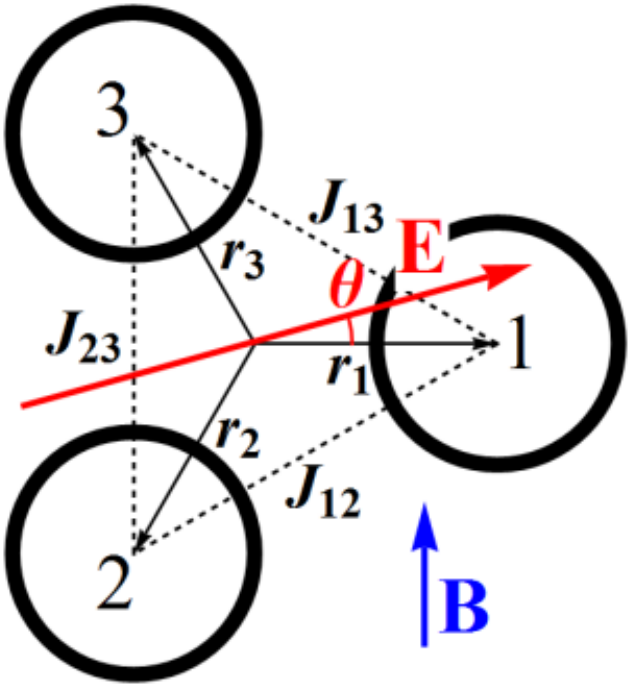}
\caption{Scheme of three quantum dots (qubits) in triangular symmetry under an effective electric filed $\mathbf{E}$ and a homogeneous magnetic field $\mathbf{B}$.}
\label{fig1}
\end{figure}

We consider a model of three quantum dots (sites) in triangular geometry (Fig.\ref{fig1}) with three electrons, and the system is under influence of a homogeneous magnetic field. The system is described by the Hubbard Hamiltonian~\cite{Bulka11,Kuba12}:
\begin{align}\label{eq:Hubbard}
    \hat{H}=&\sum_{i,\sigma}\tilde\epsilon_{i}n_{i\sigma}+
    t\sum_{i\neq j,\sigma}(c_{i \sigma}^{\dag}c_{j \sigma}+h.c.)
    +U \sum_{i}n_{i \uparrow} n_{i\downarrow}\nonumber \\&-g \mu_B  B_z\sum_{i,\sigma}{\sigma n_{i\sigma}}.
\end{align}
We assume that at each quantum dot a single energy level $\epsilon_i$ is accessible for electrons. In an experiment the position of $\epsilon_i$ can be shifted in a fully controllable way by potentials applied to local gates. Since the symmetry of the system is essential in our considerations, it is more suitable to introduce an effective electric field $\mathbf{E}$ and express $\epsilon_i\equiv\tilde{\epsilon_0}+e \mathbf{E}\cdot \mathbf{r}_i$ (see Fig.1). Here, $e$ denotes the electron charge and $\mathbf{r}_i$ is the vector to the $i$-quantum dot. Later we will use $\tilde{\epsilon_0}\equiv\epsilon_1 + \epsilon_2+ \epsilon_3=0$ and express the polarization term as $\tilde{\epsilon_i}\equiv e \mathbf{E}\cdot \mathbf{r}_i= g_E \cos[\theta+(i-1)2\pi/3]$, $g_E=e E r$, $r_i=r$ - a length of $\mathbf{r}_i$, $\theta$ - an angle between the electric field $\mathbf{E}$ and $\mathbf{r}_1$.
The second term in Eq.~(\ref{eq:Hubbard}) describes the electron hopping between the nearest quantum dots. To make our considerations more transparent we assume that hopping is the same $t_{ij}=t$ for each pair $\{ij\}$. The third term is the onsite Coulomb interaction of electrons with opposite spins on the quantum dots. The last term is the spin interaction with an external magnetic field - the Zeeman term.
Here $\mu_B$ denotes the Bohr magneton, $g$ is the electron $g\textrm{-factor}$ and the magnetic field $B_z$ is taken to be in the plane of the system - along the z-axis. (The magnetic field oriented perpendicular to the plane can lead to circulation of spin supercurrents which make the bipartite concurrence uniform \cite{Kuba12}. This aspect will not be considered in the paper.)

If the onsite Coulomb interaction $U$ is much greater than the other parameters $t$ and $g_E$, we can operate in the space of singly occupied states transforming the Hubbard Hamiltonian to an effective Heisenberg Hamiltonian
\begin{equation}\label{eq:Heis}
    \hat{H}_{eff} = \sum_{i<j}{J_{i j}\left(\mathbf{S_i} \cdot \mathbf{S_j}-\frac{1}{4}\right)}-h \sum_{i}{S_{z_{i}}}\,,
\end{equation}
where $h = g \mu_B  B_z$. The first term describes the superexchange coupling \cite{Anderson1964} (see also~\cite{Stohr2006}) between spins with the parameter $J_{ij}$ calculated to the third order in $t/U$~\cite{Kostyrko}:
\begin{equation}\label{eq:Jij}
    J_{ij}=\frac{4 t^2}{U}+\frac{4 t^2 (\tilde{\epsilon}_j-\tilde{\epsilon}_i)^2}{U^3}+\frac{8 t^3 (2 \tilde{\epsilon}_m-\tilde{\epsilon}_i-\tilde{\epsilon}_j)}{U^3},
\end{equation}
where $i,j,m$ are 3 different indices of the quantum dots and $\tilde{\epsilon}_i$ is the local single electron level shifted by the electric field. (The parameters $J_{ij}$ can be derived also for a general case with different hoppings $t_{ij}$ -- see \cite{Kostyrko}.)
The electric field is responsible for symmetry breaking in the system varying the superexchange parameters $J_{ij}$.

The states for three spins can be constructed from the two spin states, singlets and triplets, by adding an electron~\cite{Pauncz,Bulka11}. As a result one gets  the quadruplet manifold  and two doublet states with the total spin $S = 3/2$ and $S = 1/2$, respectively.
The quadruplet states with $S_z=+3/2, +1/2$ are:
\begin{align}
    \label{eq:Quad32}
    \ket{Q_{3/2}} =& \ket{\up \up \up}\,, \\
    \label{eq:Quad12}
    \ket{Q_{1/2}} =& \frac{1}{\sqrt{3}} \left(  \ket{\dn \up \up} + \ket{\up \dn \up} + \ket{\up \up \dn} \right)\,,
\end{align}
and similarly the states $\ket{Q_{-3/2}}$, $\ket{Q_{-1/2}}$ with the opposite spin direction. The quadruplet states with $S_z = \pm 1/2$ are the W states.
The doublet states $(^{1/2}D_{S_{z}})$ with $S_{z}=\pm 1/2$ can be constructed from the base states:
\begin{align}
    \label{eq:Doublet1}
    \ket{D_{1/2}}_1 =&\frac{1}{\sqrt{2}}\left( \ket{\up \up \dn} - \ket{\up \dn \up}\right)\,, \\
    \label{eq:Doublet2}
     \ket{D_{1/2}}_2 =&\frac{1}{\sqrt{6}}\left(2 \ket{\dn \up \up} - \ket{\up \up \dn} -\ket{\up \dn \up}\right)\,,
\end{align}
and similarly the states $\ket{D_{-1/2}}_1$, $\ket{D_{-1/2}}_2$ with the opposite spin direction. Notice that $\ket{D_{S_{z}}}_{1}$ is constructed from the singlet on the sites $\{23\}$ and adding an electron to the site 1, while $\ket{D_{S_{z}}}_{2}$ is built from triplets on the sites $\{23\}$ and adding an electron to the site 1.
These base states constitute a doublet subspace used in DiVincenzo's scheme \cite{DiVincenzo2000} for quantum computation based on three quantum semiconducting dot qubits (see also \cite{Bacon03}). All four doublet eigenstates in the system can be expressed as:
\begin{equation}\label{eq:eigenvecD}
          \ket{ D_{S_{z}}^{\pm} } = \cos \varphi^{\pm}\;  \ket{D_{S_{z}}}_{1} + \sin \varphi^{\pm}\;\ket{D_{S_{z}}}_{2}\,.
    \end{equation}
The phase $\varphi^{\pm}$ is given by the relation:
\begin{equation}\label{eq:phi}
\cot \varphi^{\pm}=\frac{J_{12}-2 J_{23}+J_{31} \pm 2\Delta}{\sqrt{3}(J_{31}-J_{12})}\,,
\end{equation}
where $\Delta$ is given by Eq.~\eqref{eq:DD}.
The symmetry of $\varphi^{\pm}$ depends on the choice of the base of doublet states Eq.~(\ref{eq:Doublet1}) and Eq.~(\ref{eq:Doublet2}).

\begin{figure}[t]
\includegraphics[scale=0.45]{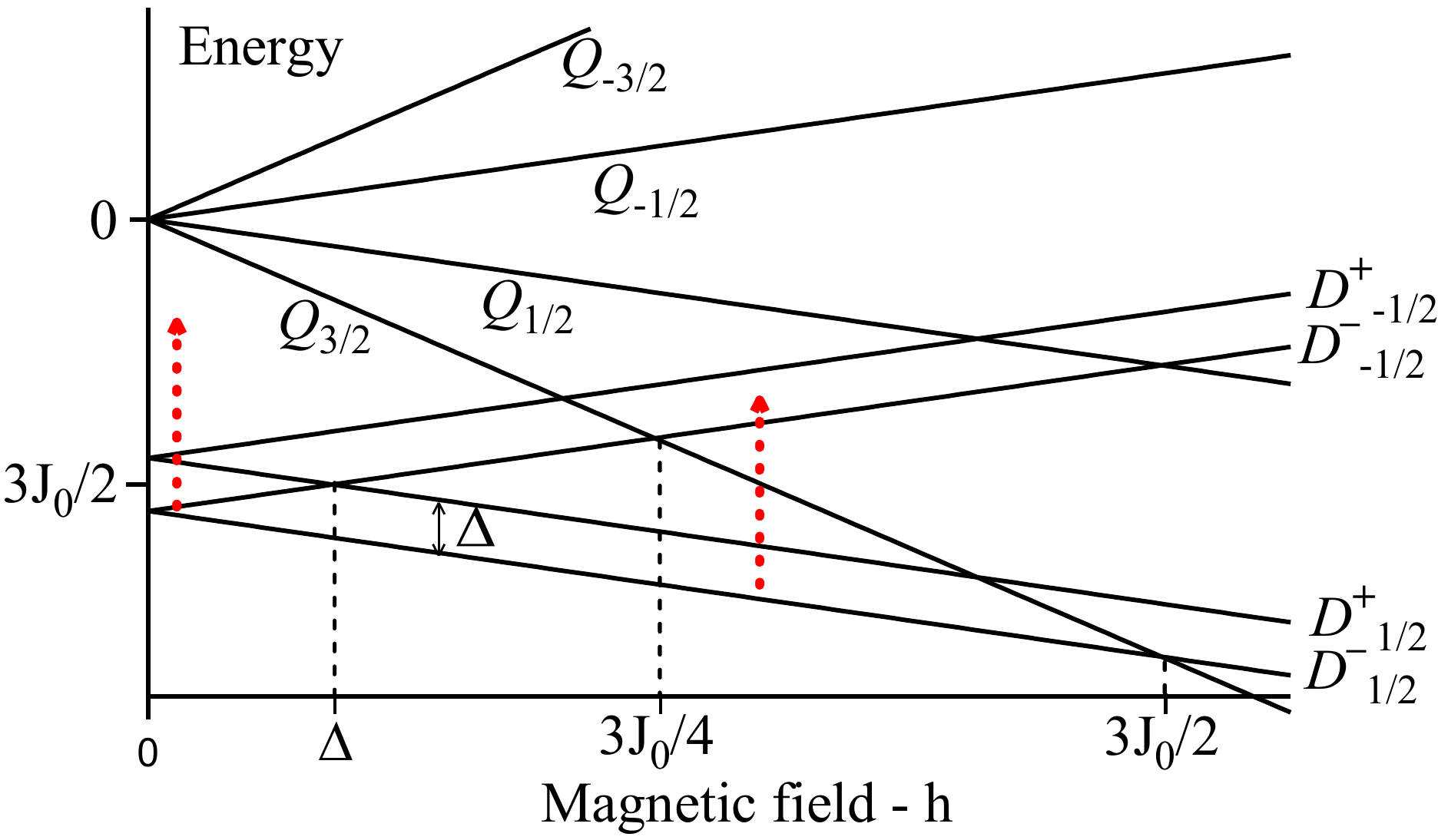}
\centering
\caption{\label{fig:energy}Scheme of energy levels with the Zeeman splitting of the doublet $(D_{\pm1/2}^{\pm})$
and the quadruplet $(Q_{\pm3/2,\pm1/2})$ states. Red dashed arrows
show two considered situations of thermal mixing of states with raising
temperature in the system. The first one is for weak magnetic field $h\ll\Delta$
and the second is for strong magnetic field ($\Delta\ll h<3J_{0}/2$).
Here $J_{0}\equiv 4t^2/U$ and we assume $\Delta\ll J_{0}$}.
\end{figure}

The existence of electric and magnetic field breaks the fourfold degeneracy
of quadruplet and doublet states in the symmetric system. Solving the eigenvalue problem for the effective Hamiltonian Eq.~(\ref{eq:Heis}) one finds the energies:
\begin{align} \label{eq:eigenvalueQ}
    E_{Q_{S_{z}}}=&- h S_z\;,\\
\label{eq:eigenvalueD}
    E_{D_{S_{z}}^{\pm}}=&-\frac{3}{2} J \pm \frac{1}{2}\Delta-h S_z\;,
\end{align}
for quadruplet and doublets, respectively. Here,
\begin{align}\label{eq:J}
   J =& (J_{12}+J_{23}+J_{31})/3,\\
   \label{eq:DD}
   \Delta=&\sqrt{J_{12}^2+J_{23}^2+J_{31}^2-J_{12}J_{31}-J_{12}J_{23}-J_{31}J_{23}}.
\end{align}
$\Delta$ is responsible for splitting of eigenstates due to the presence of electric field. Therefore we call this effect as the {\it spin Stark effect}~\cite{Kuba12}. A scheme of energy levels split by the magnetic field $h$ is shown in Fig. \ref{fig:energy}.

Experimentally it is possible to obtain similar system of three quantum
dots \cite{Seo2013} but in reality it will never be perfectly symmetrical resulting
in different hopping integrals $t_{ij}$ and different $U_{i}$ on
each dot. However if the canonical transformation is uses for the general case~\cite{Kostyrko} one can nevertheless get any desired value of $J_{ij}$ by modifying local dot potentials $\tilde{\epsilon}_{i}$ .

Notice that there are no processes for transition between the subspaces with the total spin $S=1/2$ and $S=3/2$ for our model (\ref{eq:Heis}). The doublet subspace is the decoherence free subspace (DFS) \cite{Lidar1998,Kempe} and entanglement encoded in this subspace is robust against decoherence with electrodes (an external bath system). This is a main advantage of the quantum computation scheme proposed by DiVincenzo {\it et al.} \cite{DiVincenzo2000}.

\section{Spin correlations and entanglement}

We assume that the considered spin system is in thermal equilibrium with the environment and is described by the density
matrix \cite{Linden2009}:
\begin{equation}\label{eq:densitymatrix}
    \rho = \sum_{\alpha=1}^8 z_{\alpha} \ket{\psi_\alpha}\bra{\psi_\alpha},
\end{equation}
where $z_{\alpha} = e^{- \beta E_{\alpha}}/Z$ are the Boltzmann factors with $Z=\sum_\alpha z_\alpha$, $\beta=1/k_{B}T$ , $E_{\alpha}$ are eigenenergies and the corresponding eigenvectors are $\ket{\psi_{\alpha}}\in\{\ket{D_{1/2}^{-}}$,
$\ket{D_{-1/2}^{-}}$, $\ket{D_{1/2}^{+}}$, $\ket{D_{-1/2}^{+}}$,
$\ket{Q_{3/2}}$, $\ket{Q_{1/2}}$, $\ket{Q_{-1/2}}$, $\ket{Q_{-3/2}}\}$.
We would like to study entanglement between spins on two sites $\{ij\}$ in the
considered three qubit system. To this end we use the
reduced density matrix $\rho_{ij}=\textrm{Tr}_{m\neq i,j}[\rho]$, which in the standard basis $\{\up\up,\up\dn,\dn\up,\dn\dn\}$ is expressed as:
\begin{equation}
\rho_{ij}=\begin{pmatrix}\mu_{ij} & 0 & 0 & 0\\
0 & \xi_{ij} & \gamma_{ij} & 0\\
0 & \gamma_{ij} & \nu_{ij} & 0\\
0 & 0 & 0 & \delta_{ij}
\end{pmatrix}.\label{eq:roij}
\end{equation}
Explicitly for the effective Heisenberg Hamiltonian Eq.~(\ref{eq:Heis}) the elements $\gamma_{ij}$, $\mu_{ij}$ and $\delta_{ij}$ are expressed as:
\begin{equation}\label{eq:roijpar}
\begin{split}
     \mu_{ij} =&\frac{1}{3}[(z_{D_{1/2}^+} -  z_{D_{1/2}^-}) \cos(\tilde{\theta}_{ij}) + z_{D_{1/2}} ]\\
    &+z_{Q_{3/2}}+\frac{z_{Q_{1/2}}}{3},\\
     \gamma_{ij} =& [(z_{D^+} -  z_{D^-} ) 2 \cos(\tilde{\theta}_{ij}) - z_D ]\\
    &+\frac{z_{Q_{1/2}} + z_{Q_{-1/2}}}{3},\\
     \delta_{ij} =& \frac{1}{3}[(z_{D_{-1/2}^+} -  z_{D_{-1/2}^-}) \cos(\tilde{\theta}_{ij}) + z_{D_{-1/2}}]\\
    &+z_{Q_{-3/2}}+\frac{z_{Q_{-1/2}}}{3},
\end{split}
\end{equation}
where
\begin{align}
    z_{D}& \equiv z_{D_{-1/2}^{-}} +z_{D_{-1/2}^{+}} +z_{D_{+1/2}^{-}} + z_{D_{+1/2}^{+}}, \nonumber \\ \label{eq:zD}
    z_{D^{\pm}}& \equiv z_{D_{-1/2}^{\pm}}+z_{D_{+1/2}^{\pm}}, \\ z_{D_{\pm1/2}}& \equiv z_{D_{\pm1/2}^{+}}+z_{D_{\pm1/2}^{-}}.\nonumber
\end{align}
In Eq.~\eqref{eq:roijpar} $\tilde{\theta}_{ij}=2\varphi^-+q_{ij}$, $\varphi^-$ is the phase given by Eq.~\eqref{eq:phi} and $q_{ij} = \{0,-2 \pi/3,2 \pi/3\}$ for site pairs $\{ij\} = \{23,13,12\}$, respectively. The form of $q_{ij}$ comes from the chosen symmetry of the base states Eq.~\eqref{eq:Doublet1}, Eq.~\eqref{eq:Doublet2}. In derivation of Eq.~\eqref{eq:roijpar} the relation: $\cos \left(2 \varphi^- + q_{ij}\right) = - \cos \left(2 \varphi^+ + q_{ij}\right)$ was used which can be proved using Eq.~(\ref{eq:phi}) and Eq.~(\ref{eq:DD}).

We use concurrence $C_{ij}$ as a measure of pair-wise entanglement \cite{Hill97}, which is given by $C_{ij} = \max\{\Lambda_{ij}=\lambda_{\mathrm{max}} - \sum_i \lambda_i,0\}$, where $\lambda_i$ are eigenvalues  of the matrix $\rho_{ij}\cdot \tilde{\rho}_{ij}$ and  $\tilde{\rho}_{ij}=(\sigma_y \bigotimes \sigma_y)\cdot \rho_{ij}^*\cdot(\sigma_y \bigotimes \sigma_y)$ is spin flipped matrix. For spin rings and the density matrix of the form Eq.~(\ref{eq:roij}) concurrence can be expressed as \cite{OConnor}:
\begin{equation}\label{eq:Cro}
    C_{ij} = 2 \max\{|\gamma_{ij}|-\sqrt{\mu_{ij} \delta_{ij}},\,0\}\,.
\end{equation}

In further investigations of thermal entanglement we use the parameters Eq.~\eqref{eq:roijpar} for the effective Heisenberg model Eq.~\eqref{eq:Heis} with superexchange couplings. The exact solutions of thermal concurrence together with the analysis of several special cases and relationships between spin correlation function and concurrence will be presented below. It is worth to notice that these results are valid for general case of any Heisenberg model, in which one can change symmetry of the spin system by a modulation of exchange couplings $J_{ij}$.

\subsection{Ground state}\label{sub:grounState}

First we present studies for entanglement in the ground state, which will be a reference system for further studies of thermal entanglement when temperature leads to mixing with excited states.  For magnetic field $0<h<3J_{0}/2$ the ground state is $\ket{D_{1/2}^{-}}=\cos\varphi^{-}\;\ket{D_{1/2}}_{1}+\sin\varphi^{-}\;\ket{D_{1/2}}_{2}$ (see Fig.\ref{fig:energy}). The symmetry of the system is controlled by orientation of the electric field, by its angle $\theta$.  Using Eq.~\eqref{eq:Cro} and Eq.~\eqref{eq:roijpar}
one can easily show that $C_{ij}$ reads:
\begin{equation}
C_{ij}  =  \frac{1}{3}\left|2\cos(\tilde{\theta}_{ij})+1\right|.\label{eq:CDum}
\end{equation}
Comparing with a spin-spin correlation function:
\begin{equation}
\mSiSj{i}{j}  =  \frac{1}{4}\left[-2\cos(\tilde{\theta}_{ij})-1\right]\label{eq:SSDum}
\end{equation}
one finds the following relation \cite{Kuba12}:
\begin{equation}
C_{ij}  =  \frac{4}{3}|\mSiSj{i}{j}|.\label{eq:CSSDum}
\end{equation}
 As a result it appears that in this case, the concurrence and the expectation
value for the spin correlation function contain the same quantum information.

\begin{figure}[t]
\centering
\includegraphics[scale=0.8]{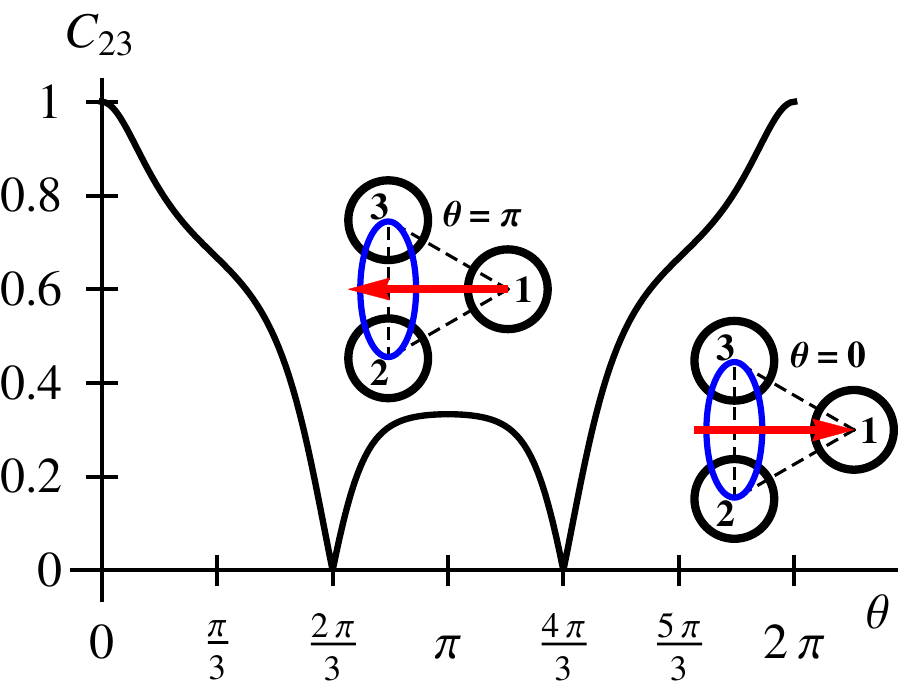}
\caption{\label{fig:CDum} Plot of concurrence $C_{23}$ for the doublet ground state $\ket{D^-_{1/2}}$ as
a function of angle $\theta$ of electric field. The insets show the
orientation of electric field (red arrow) in relation to the considered
pair for $\theta=0$ and $\theta=\pi$, when concurrence becomes maximal $C_{23}=1$ and $C_{23}=1/3$, respectively. Parameters used for computation: $U=20$, $g_E=1$, $t=1$.}
\end{figure}

The plot of concurrence $C_{23}$ Eq.~\eqref{eq:CDum} as a function of $\theta$ is shown in Fig. \ref{fig:CDum}. Here we choose the pair $\{2,3\}$ because it is most convenient for the doublet basis Eq.~(\ref{eq:Doublet1})-Eq.~(\ref{eq:Doublet2}). Due to the symmetry one can obtain identical plots for $C_{12}$ and $C_{13}$ shifting the plot for $C_{23}$ by $\theta=\pm2\pi/3$.
The parameters for numerical computations were taken: Coulomb repulsion $U=20$, hopping integral $t=1$ and electric field strength $g_E=1$, which ensure the requirement of $U\gg t, g_E$ for the canonical transformation to Heisenberg Hamiltonian. For this strength of the electric field the parameter $\tilde{\theta}_{23}\approx \theta$.
One can see that the concurrence reaches maximum $C_{23}=1$ when the  electric field is perpendicular to the considered pair and oriented towards the site $1$.
Then the local electron energy is minimal and the coupling $J_{23}$ is maximal, $J_{23}>J_{12}=J_{13}$. In this case the ground state of the system becomes $\ket{D_{1/2}}_{1}$
Eq.~\eqref{eq:Doublet1} which describes singlet on the pair $\{23\}$
and separated spin at the site $1$. This means monogamy, because the concurrence between the other sites, $C_{12}=C_{13}=0$ \cite{Coffman}. Two minima $C_{23}=0$ at $\theta= 2\pi/3$ and $4\pi/3$
correspond to singlets on pairs $\{12\}$ and $\{13\}$, respectively.
Additionally one can see local maximum of $C_{ij}=1/3$ at $\theta=\pi$ corresponding
to the situation when the local electron energy is maximal and the coupling $J_{23}$ is minimal, $J_{23}<J_{12}=J_{13}$, with the ground state $\ket{D_{1/2}}_{2}$, Eq.~\eqref{eq:Doublet2}, and the correlation between the spins is ferromagnetic, $\mSiSj{2}{3}=+1/4$. Notice that switching the orientation of electric field from $\theta=0$ to $\theta=\pi$ causes a rotation of DiVincenzo's qubit from the state $\ket{D_{1/2}}_{1}$ to $\ket{D_{1/2}}_{2}$.
In the following we will investigate how these features survive thermal mixing with excited states.

\subsection{Mixed states}

For finite temperature $T>0$ we assume that the system is in thermal equilibrium and the states are populated according the Boltzmann factor $z_{\alpha}$. Note that for a non-degenerate ground state the thermal entanglement at $T=0$ is identical to the entanglement of the ground state considered in previous subsection. The arrangement of eigenenergies $E_{\alpha}$ in this case mainly depends on the magnetic field $h$ (see Fig. \ref{fig:energy}). On the other hand the doublet states $\ket{D_{\pm1/2}^{\pm}}$ Eq.~\eqref{eq:eigenvecD} depend on the angle of electric field $\theta$ which modulates parameters $\varphi^{\pm}$ and corresponding $\tilde{\theta}_{ij}$.
We would like to show how the concurrence $C_{ij}(\theta)$ for the ground state Eq.~\eqref{eq:CDum} changes with an increase of temperature $T$, resulting in mixing with excited states, for different magnetic fields $h$. In particular, two cases of weak and strong magnetic field will be considered, as they correspond to different order of eigenstates that qualitatively change system thermal properties and entanglement.

Using Eq.~\eqref{eq:Cro} and Eq.~\eqref{eq:roijpar} we derive a general formula for the concurrence taking into account mixing with all states:
\begin{align} \label{eq:CijAll}
    C_{ij}=&\frac{1}{3}\Big[\big|z_{D}-2(z_{Q_{-1/2}}+z_{Q_{+1/2}})-2(z_{D^{+}}-
    z_{D^{-}})  \nonumber \\
    &\times\cos(\tilde{\theta}_{ij})\big|
    -2\sqrt{K_{ij}^{\dn}K_{ij}^{\up}}\,\Big]\,,
\end{align}
where
 \begin{align} \label{eq:CijAllK}
    K_{ij}^{\up} =&  z_{D_{+1/2}}+z_{Q_{+1/2}}+3z_{Q_{+3/2}} \nonumber \\
    &-(z_{D_{+1/2}^{-}}-z_{D_{+1/2}^{+}})\cos(\tilde{\theta}_{ij})
\end{align}
and similarly $K_{ij}^{\dn}$ with opposite spin direction. Here the Boltzmann coefficients are described by Eq.~\eqref{eq:zD}.
Because the electric field acts only on doublets states, the concurrence Eq.~(\ref{eq:CijAll}) becomes independent on $\theta$ when the doublets $\ket{D^-_{S_z}}$ and $\ket{D^+_{S_z}}$ are thermally mixed, {\it i.e.}, for $z_{D^{-}}\approx z_{D^{+}}$.

\begin{figure}[t]
\centering
  \includegraphics[width=0.4\textwidth]{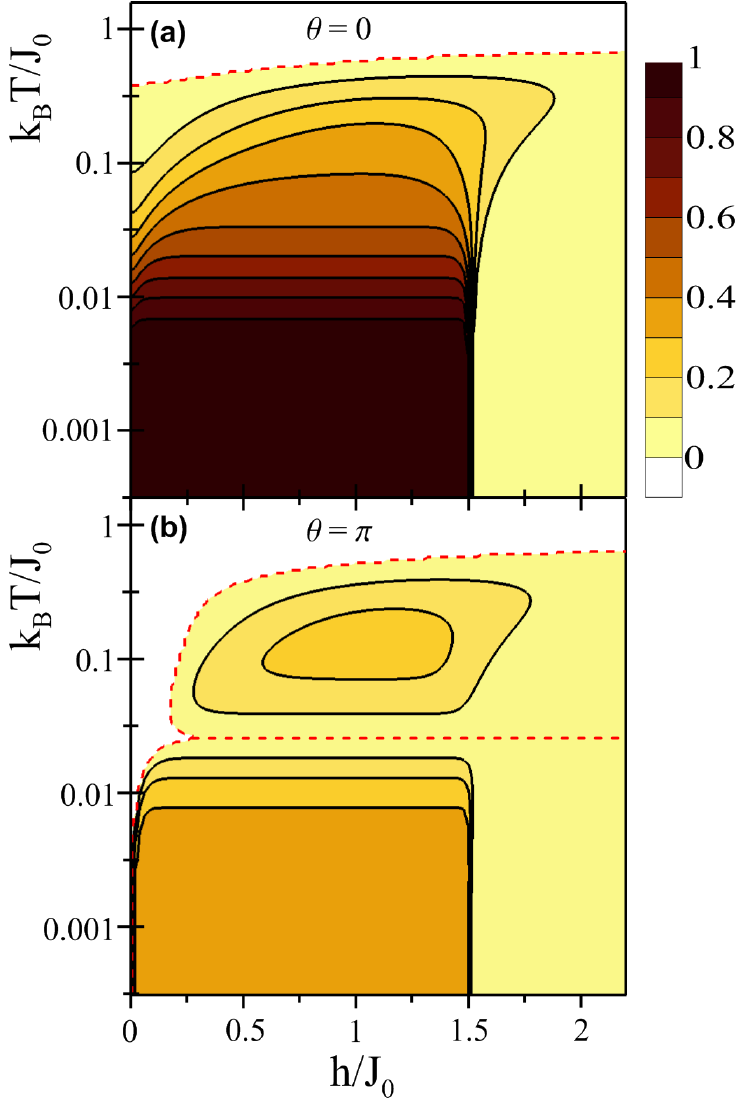}
  \caption{\label{fig:cht} Contour plot for concurrence $C_{23}$ in the $h$-$T$ plane for an angle (a) $\theta=0$ and (b) $\theta=\pi$  of electric field. The dashed (red) curve corresponds to $T_0$ when $C_{23}=0$ and the white area shows the unentangled region with $C_{12}=0$. Parameters used for computation: $U=20$, $g_E=1$, $t=1$, which give $J_0\equiv 4t^2/U=0.2$, $\Delta \approx 0.006=0.03 J_0$.}
\end{figure}

\begin{figure}[t]
\centering
  \includegraphics[width=0.4\textwidth]{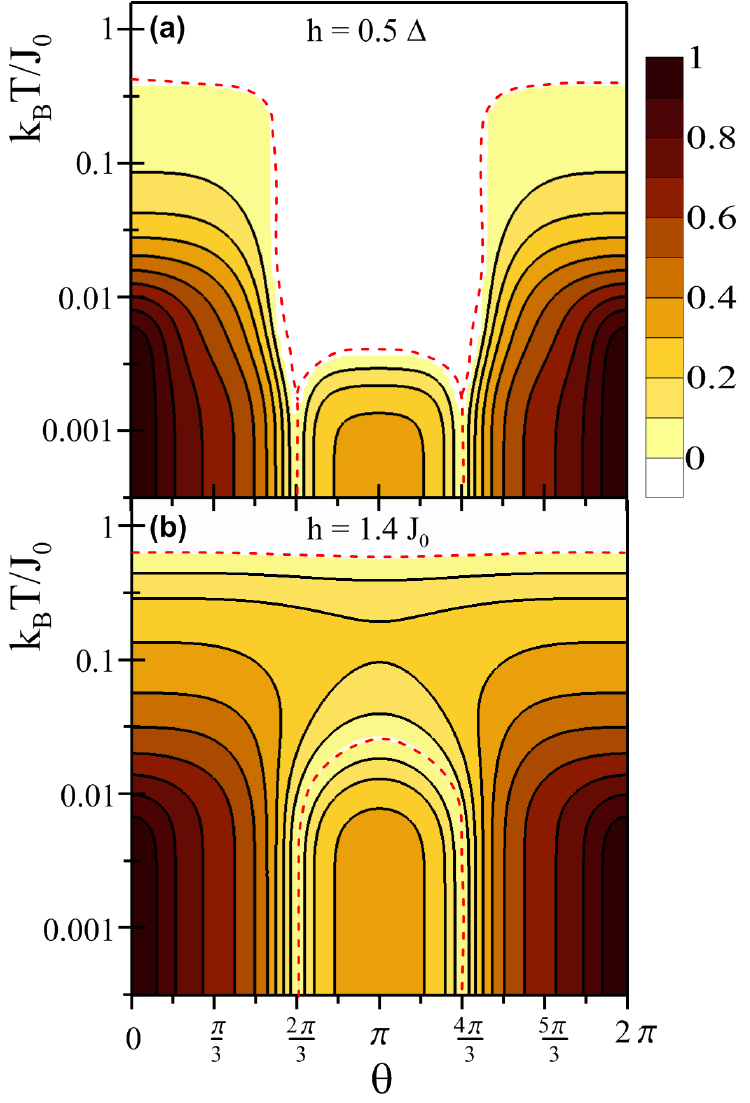}
  \caption{\label{fig:cthetat} Contour plot for concurrence $C_{23}$ in the $\theta$-$T$  plane (a) for a weak {[}$h=0.5\Delta\approx0.015J_{0}${]}, and (b) strong {[}$h=1.4J_{0}${]} magnetic field. The two magnetic field regimes corresponds to dashed arrows on energy plot \ref{fig:energy}. Parameters are the same as in Fig.\ref{fig:cht}.}
\end{figure}

Dependence of the concurrence on magnetic field is presented on the $h$-$T$ plots in Fig.~\ref{fig:cht} for the case $\theta=0$ and $\pi$, which correspond to the electric field $\mathbf{E}$ perpendicular to the bond $\{23\}$ and directed from and to the site $1$, respectively.
In the absence of the magnetic field ($h=0$) the states with opposite spin orientations are degenerated and equally populated. In the ground state the concurrence is given by $C_{ij}(T=0,h=0)=\max\{\cos(\tilde{\theta}_{ij}),\,0\}$. Switching on the magnetic field removes degeneracy and results in a large increase of concurrence up to $h<\Delta$. For higher magnetic fields the doublets $\ket{D^-_{1/2}}$ and $\ket{D^+_{1/2}}$ with the same spin orientations play a crucial role in low temperatures. Concurrence $C_{ij}(T)$ increases with $h$, because the larger Zeeman splitting makes weaker thermal mixing.
For a very large magnetic field, $h>3J_0/2$,  the unentangled quadruplet  $Q_{3/2}$ is the ground state and hence the concurrence diminishes for very strong magnetic fields \cite{Arnesen,Wang01}. In this range the concurrence is exponentially small due to thermal excitations to the doublets.

An interesting result is presented in Fig.~\ref{fig:cht}(b) where the concurrence shows a nonmonotonic temperature dependence, first at low temperatures it decreases to zero and one observes restoration of entanglement at high temperatures.  We show below that this is an effect of  a specific interplay between entanglement of the ground state $\ket{D^-_{1/2}}$ and the excited state $\ket{D^+_{1/2}}$.

Fig. \ref{fig:cthetat} shows contour plots of $C_{23}$ in the plane of  $\theta$  and $T$ for two different values of magnetic field $h$, weak ($h<\Delta$) and strong ($\Delta \ll h < 3J_0/2$).
One can see two local maxima $C_{23}=1$ and $C_{23}= 1/3$ for the angles $\theta=0$ and $\theta=\pi$, as well as two minima with $C_{23}=0$ at $\theta=2\pi/3$ and $4\pi/3$. For weak magnetic field $C_{23}(T)$ monotonously decreases with an increase of $T$, while for high magnetic field $C_{23}(T)$ is a nonmonotonic function at $\theta=\pi$ but it is still monotonic at $\theta=0$. Let us analyze thermal entanglement in these two cases in details.

\paragraph{Weak magnetic field}

At low temperatures one can consider mixing of two states
only: the ground state $\ket{D_{1/2}^{-}}$.
These states have the same symmetry (with the same phase $\varphi^{-}$) but opposite spin orientation.
In this case one can take into account only the Boltzmann coefficients $z_{D_{1/2}^{-}}$ and $z_{D_{-1/2}^{-}}$
in Eq.~(\ref{eq:CijAll}) and write the concurrence as:
\begin{align}
C_{ij}=&\max\big\{ \frac{1}{3}\left|2\cos(\tilde{\theta}_{ij})+1\right| \nonumber \\
&-\frac{2}{3}\sqrt{z_{D_{1/2}^{-}}z_{D_{-1/2}^{-}}}\left[1-\cos(\tilde{\theta}_{ij})\right],\,0\big\} \,.\label{eq:CDudm}
\end{align}
Comparing this result with the concurrence Eq.~\eqref{eq:CDum} for the ground state
one can see that the second term in Eq.~\eqref{eq:CDudm} describes reduction
of the concurrence due to thermal mixing of the states. Notice that for $\cos(\tilde{\theta}_{ij})=1$ the mixing term vanishes and $C_{ij}=1$ is sustained. It means that thermal entanglement is resistant to temperature rise till the higher energy states ($\ket{D_{\pm1/2}^{+}}$) become populated.

The reduction term is proportional to  $[z_{D_{1/2}^-} z_{D_{-1/2}^-}]^{1/2}\approx  e^{-\beta h/2}$. This result may be compared to the one obtained by Gunlycke {\it et al.} \cite{Gunlycke} for the Ising model in a perpendicular magnetic field, where the thermal mixture of two pure qubit states cause the reduction of concurrence proportional to $e^{-\beta h}$. Such a result is valid for any thermal mixing of pure states $\ket{\psi_n}$ with no spin-flip overlap, i.e. for $\bra{\psi_n}\sigma_y \bigotimes \sigma_y\ket{\psi_m}=0$. In our case, however, spin-flip overlap of mixed stated does exist and the reduction of entanglement is lower, as $e^{-\beta h/2}$.

At high temperatures $k_{B}T\gg h$ all states are populated and one
may assume total mixing between the states with the opposite spins
i.e.:  $z_{D_{1/2}^{-}}=z_{D_{-1/2}^{-}}=z_-$,
$z_{D_{1/2}^{+}}=z_{D_{-1/2}^{+}}=z_+$ as well as $z_{Q_{\pm3/2}}=z_{Q_{\pm1/2}}=z_Q$. Then a mixed state of the system can be written in the form: $\rho=z_- \rho^- + z_+ \rho^+ + z_Q \rho^Q$, where $\rho^{\pm}=\sum_{S_z}\ket{D_{S_z}^{\pm}}\bra{D_{S_z}^{\pm}}$ and $\rho^{Q}=\sum_{S_z}\ket{Q_{S_z}}\bra{Q_{S_Z}}$. Partial concurrences of those constituent states: $C_{ij}(\rho^{\pm,Q})=\max\{\Lambda_{ij}^{\pm,Q},\,0\}$ can be used to express concurrence $C_{ij}(\rho)$:
\begin{equation}\label{CeqAll2}
    C_{ij}=\max\{ z_+ \Lambda_{ij}^{+}+ z_- \Lambda_{ij}^{-}+3 z_Q \Lambda_{ij}^{Q},\,0 \}\,,
\end{equation}
where $\Lambda_{ij}^{\pm,Q}=2\left(|\gamma_{ij}|-|\mu_{ij}|\right)$ Eq.~\eqref{eq:Cro}. It shows that concurrence is expressed as thermal redistribution with additional factor $3$ for quadruplet states. Using Eq.~\eqref{eq:roijpar} we get
\begin{equation}
C_{ij}=\max\{(z_{-}-z_{+})\cos(\tilde{\theta}_{ij})-z_{Q},\,0\}\;.\label{eq:CeqAll}
\end{equation}

\paragraph{Strong magnetic field}

For the case of strong magnetic field ($\Delta<h<3J_{0}/2$) we have at low temperatures a mixture of two lowest
states $D_{1/2}^{-}$ and $D_{1/2}^{+}$ with the same spin orientation. The symmetry of these states is different
and characterized by different phases $\varphi^{-}$ and  $\varphi^{+}$. In this limit the concurrence Eq.~\eqref{eq:CijAll}
can be expressed as
\begin{equation}
C_{ij}=\frac{1}{3}\left|(z_{D_{1/2}^{-}}-z_{D_{1/2}^{+}})2\cos(\tilde{\theta}_{ij}) +(z_{D_{1/2}^{-}}+z_{D_{1/2}^{+}})\right|.\label{eq:CDupm}
\end{equation}
Here the dependence on $\tilde{\theta}_{ij}$ exhibits specific interplay between entanglement in the doublet states. For $\cos(\tilde{\theta}_{ij})>0$ (i.e. for $\theta \in [0,2\pi/3]$ and $\theta \in [4\pi/3,2\pi]$ in the case presented in Fig.~\ref{fig:cthetat}) the concurrence decreases monotonously with increasing $T$.
On the other hand, if $\cos(\tilde{\theta}_{ij})<0$ (for $\theta \in [2\pi/3,4\pi/3]$ in Fig.~\ref{fig:cthetat}), the entanglement dominated by the excited state is larger than the one of the ground state. In this range the expectation value of the spin correlation function $\mSiSj{i}{j}_{D^-_{1/2}}>0$ for the state $\ket{D^-_{1/2}}$, whereas the expectation value $\mSiSj{i}{j}_{D^+_{1/2}}<0$ for the state $\ket{D^+_{1/2}}$. The specific interplay between entanglement of both states leads to a reduction of $C_{ij}$ to zero for small $T$, and its reconstruction for higher $T$ as entanglement coming from the excited state becomes dominant. One can calculate the critical temperature $k_BT_0= \Delta/\ln 3$ at which $C_{ij}=0$ due to perfect compensation of the both contributions from the state $D_{1/2}^-$ and $D_{1/2}^+$.

Similar situation when entanglement is restored one can observe  for very strong magnetic field $h>3J_{0}/2$ with quadruplet $\ket{Q_{3/2}}$  as the ground state \cite{Arnesen,Wang01}. Quadruplet is unentangled state but for higher $T$ doublets become populated which results restoration of concurrence.

\paragraph{Spin-spin correlation functions}

One can also find the formula for spin-spin correlation functions
for the mixed state Eq.~\eqref{eq:Cro} which reads:
\begin{equation}
\mSiSj{i}{j}=\frac{1}{4}(-z_{D}+z_{Q}+2(z_{D^{+}}-z_{D^{-}})
\cos(\tilde{\theta}_{ij})).\label{eq:SiSjAll}
\end{equation}
Comparing general formula for concurrence Eq.~\eqref{eq:CijAll} and spin-spin correlation functions Eq.~\eqref{eq:SiSjAll} one can see that
it is not possible to simply relate them in a general case.
However, at small temperature $T$ and magnetic field $\Delta<h<3J_{0}/2$ only doublets $\ket{D_{+1/2}^{\pm}}$ with spin up are populated (see the energy plot in Fig. \ref{fig:energy}) and the relation between $C_{ij}$
and $\mSiSj{i}{j}$ can be expressed as: $C_{ij}  =  \frac{4}{3}|\mSiSj{i}{j}|$ as for the case of ground state Eq.~\eqref{eq:CSSDum}. On the other hand, in the case of very small $h \ll T$ when the Zeeman spitting is negligible and the states with opposite spin direction are equally populated, the relation changes to $C_{ij}=\max\left\{ -2\mSiSj{i}{j}-\frac{1}{2},\,0\right\} $\cite{Ramsak09}.
In the general case the relation between $C_{ij}$ and $\mSiSj{i}{j}$
lies somewhere between those two extrema
\begin{equation}
\max\{ -2\mSiSj{i}{j}-\frac{1}{2},\,0\} \leq C_{ij}\leq\frac{4}{3}\left|\mSiSj{i}{j}\right|.\label{eq:CSSGeneral}
\end{equation}

\section{Summary}

Summarizing, we have investigated thermal pairwise entanglement in a triangular system of three coherently coupled semiconducting quantum dots. Spin correlations have been described within the effective Heisenberg Hamiltonian in which exchange coupling constants are derived from the Hubbard model using the canonical perturbation theory and tracing out double occupied states. The investigations included the Zeeman splitting caused by the magnetic field as well as symmetry breaking by the electric field (the spin Stark effect). We have shown that spin entanglement is generated by the doublet states. Rotating the electric field one can change the entanglement, creating
maximum entanglement in a chosen pair of qubits together with a separate uncoupled spin (a spin dark state).
For a specific symmetry one can set the system in the ground state either in state $\ket{D_{1/2}}_{1}$ (for $\theta=0$) or in state $\ket{D_{1/2}}_{2}$ (for $\theta= \pi$) and by  manipulating the orientation $\theta$ of the electric field one can easily prepare the qubit in a desired state on the Bloch sphere. Our studies of entanglement indicate that the poles on the Bloch sphere (corresponding for $\ket{D_{1/2}}_{1}$ and $\ket{D_{1/2}}_{2}$) are the most stable and easiest to prepare (see Fig.\ref{fig:CDum}).

In small clusters of coupled electrons with strong Coulomb repulsion it is not uncommon to find states corresponding to perfectly entangled qubit pairs. However, in general such states can be unstable with respect to charge fluctuations, coupling to energetically near states excited due to elevated temperature, small external magnetic fields or due to the coupling to external charge reservoirs. In this paper we concentrated to the properties of charge-transfer isolated TQD, but in contact with thermal bath in the presence of an external magnetic field, while the analysis of the influence of the charge-transfer coupling to external leads will be presented elsewhere.

As expected, at finite temperatures the entanglement will in general be reduced due to thermal mixing with excited states. In particular, our studies show that the state $\ket{D_{1/2}}_{1}$ is robust on thermal mixing and stable for relatively large temperatures (see Fig. \ref{fig:cht} and \ref{fig:cthetat}). In contrast the state $\ket{D_{1/2}}_{2}$ is less stable on temperature. For a high temperature the excited states (with different pairwise correlations between spins) will come into the play.
Concurrence depends on relative thermal occupation of both states and it can be a non-monotonic function of temperature. At the regime of very low temperatures and not too large magnetic field (the ground state is doublet), thermal mixing reduces the concurrence. However, at higher temperatures, the entanglement can be restored due to spin correlations in the excited state. As expected, thermal mixing between the doublets with opposite spin orientation, $S_z=+1/2$ and $S_z=-1/2$, reduces the entanglement. Quadruplets (the unentangled states) lead to total destruction of entanglement.

\section*{Acknowledgments:}
This work was supported by the National Science Centre under the contract  DEC-2012/05/B/ST3 /03208 and by the EU project Marie Curie ITN NanoCTM. A.R. would like to acknowledge also support from ARRS under contract no. P1-0044.

\bibliography{references}   

\end{document}